\documentclass[reprint,twocolumn,superscriptaddress,longbibliography]{revtex4-2}

\usepackage{color}

\definecolor{nblue}{RGB}{28,130,185}

\usepackage[normalem]{ulem}

\pdfoutput=1

\usepackage{tikz}
\usepackage{hyperref}
\hypersetup{
  colorlinks=true,
  citecolor=blue,
  urlcolor=blue,
  linkcolor=blue
}

\usepackage{mathtools}
\usepackage{amssymb}

\definecolor{darkgreen}{rgb}{0,.4,0}
\definecolor{mixedgreen}{rgb}{0.3,0.6,00}

\newcommand{\REMOVE}[1]%
           {{\color{magenta}\sout{#1}}}

\begin{document}

\title{ Absorbing phase transitions with memory in critical scaling}
\author{Kartik Chhajed}
\thanks{kartik@pks.mpg.de}
\affiliation{Max-Planck Institute for Physics of Complex Systems,  Nöthnitzer Str. 38, 01187 Dresden, Germany}
\author{P. K. Mohanty} 
\thanks{pkmohanty@iiserkol.ac.in}
\affiliation{Max-Planck Institute for Physics of Complex Systems,  Nöthnitzer Str. 38, 01187 Dresden, Germany}
\affiliation{Department of Physical Sciences, Indian Institute of Science Education and Research Kolkata, Mohanpur, 741246 India}

\begin{abstract}
Many driven systems alternate between bursts of activity and quiescence and can become trapped in an absorbing state, such as complete inactivity in reaction–diffusion processes or extinction in predator–prey dynamics. It is generally assumed that, conditioned on survival, their long-lived (quasi-stationary) behavior is unique and independent of the initial condition. We show this need not hold, even for memoryless Markov dynamics. When the  configuration space fractures into multiple macroscopic communicating classes, where configurations  can be reach from one another  within a class but not across classes, the system retains a measurable memory of its preparation, which can directly affect the critical exponents near absorbing transitions. Using a minimal birth–death–diffusion model, we demonstrate that the quasi-stationary state is unique when birth processes are present, but becomes nonunique and initial-condition dependent when they are suppressed. This mechanism, arising from vanishing of inter-class escape-rate ratios in thermodynamic limit, challenges the conventional universality hypothesis and suggests possibility of history‑dependent critical scaling in controlled lattice or colloidal systems with  tunable particle-number. 
\end{abstract}

\maketitle

\section{Introduction}
Markov processes are used extensively in physics, chemistry, biology, economics, and many other fields to model real-world dynamics under both equilibrium and nonequilibrium conditions \cite{Kampen2007,Gillespie1992}. Lattice models in thermal equilibrium \cite{Baxter1982,Friedli2017,Lavis2015} are typically described by Markov-chain Monte Carlo dynamics that satisfy detailed balance with respect to a Gibbs measure; such approaches have found wide application in polymer physics \cite{Vanderzande1998}, active and granular media \cite{Manacorda2018}, traffic flows \cite{Schadschneider2010}, and protein folding \cite{Abeln2014}. In the absence of a Gibbs measure, nonequilibrium systems are commonly modeled by continuous-time Markov jump processes \cite{Privman1997}, which display genuinely nonequilibrium phases and phase transitions \cite{Dickman1999,Henkel2008}.

Ergodicity is essential for the emergence of a unique steady state. This condition is satisfied when the Markov chain is irreducible, i.e., when every configuration can be reached from any other configuration in a finite number of steps. Corresponding Markov matrices are irreducible and the uniqueness of its largest eigenvalue and the uniqueness of the stationary state are well protected by the Perron–Frobenius (PF) theorem \cite{Keizer1972,Berman1994,Seneta2006}. 

\begin{figure}[!h]
  \includegraphics[width=\columnwidth]{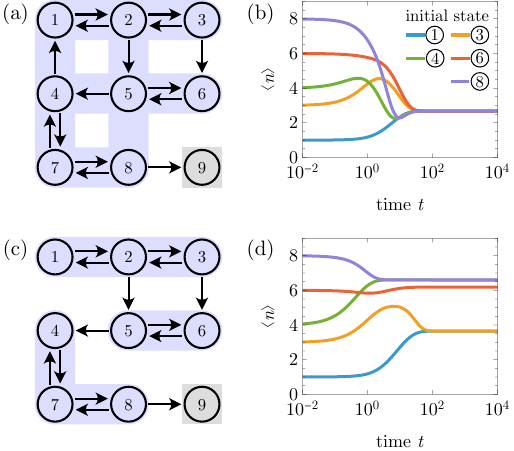}
  \caption{
    \textbf{Communicating class (CC) structure controls the uniqueness of long‑time survival‑state behavior.} Nodes are states \(n\in S=\{1,\dots,9\}\); arrows indicate allowed transitions. Open CCs: rounded blue; closed CCs: grey. 
    \textbf{(a)} One open CC \(S\setminus\{9\}\) and one closed CC \(\{9\}\). 
    \textbf{(b)} Time evolution of the average survival state \(\langle n(t)\rangle\) (conditional on not being in the closed CC), started from \(n_i\in\{1,3,4,6,8\}\) and computed for the network in (a); all curves converge to the same asymptotic value, reflecting unique long‑time behavior. 
    \textbf{(c)} Multiple open CCs \(\{1,2,3\}\), \(\{5,6\}\), \(\{4,7,8\}\) and one closed CC \(\{8,9\}\). 
    \textbf{(d)} \(\langle n(t)\rangle\) for the network in (c), initialized at the same \(n_i\); here the asymptotic form depends on the initial open CC, indicating memory of preparation.}
  \label{Fig:FigExplainCC_alternative}
\end{figure}

Ergodicity is often taken for granted, but it is trivially broken in systems having absorbing configurations, which can be reached by the dynamics but, once reached, trap the system indefinitely. The long-time behaviour of such systems is governed by the open communicating classes defined as follows. In a Markov chain, two configurations belong to the same communicating class (CC) if each is accessible from the other. A “closed” CC is one that cannot be left; there is no path that exits the class. CCs that are not closed are “open”; at least one path exits the class (for formal definitions, see Appendix~\ref{AppendixSec:CommunicatingClasses} or \cite{Norris1998,Levin2017}).
Open and closed communicating classes (CCs) are illustrated in Fig.~\ref{Fig:FigExplainCC_alternative} using two examples of Markov dynamics on the state space $S=\{0,1,\dots,9\}$.

In Fig. \ref{Fig:FigExplainCC_alternative}(a) the network consists of a single open communicating class (CC) and a single closed CC. When the nonabsorbing part of the network is irreducible, the Perron–Frobenius theorem implies a unique long‑time behavior. Accordingly, the average survival state \(\langle n(t)\rangle\), with \(n\in S=\{1,2,\dots,9\}\) the node index shown in the graphs, converges to the same long‑time form for all tested initial states \(n_i\in\{1,3,4,6,8\}\) [Fig. \ref{Fig:FigExplainCC_alternative}(b), computed for the network in (a)]. See Appendix~\ref{AppendixSec:MultipleOpenCCs} for explicit choices of transition rates. The network in panel (c) is obtained by removing the transition \(4\to 1\) from panel (a), which splits the nonabsorbing part into multiple open CCs while keeping a single closed CC. In this reducible case, the long‑time behavior of \(\langle n(t)\rangle\) depends on which open CC the process is initialized in [Fig. \ref{Fig:FigExplainCC_alternative}(d), computed for the network in (c)]: trajectories started in different open CCs retain distinct asymptotic forms. This example motivates the central questions of our manuscript:
\begin{itemize}
  \item Do thermodynamic Markov systems whose active configuration space fractures into multiple macroscopic open CCs generically exhibit nonunique long time behavior, or are additional dynamical criteria required?
  \item If such systems undergo a phase transition, can nonuniqueness imprint on critical scaling, e.g., making critical exponents depend on the initial condition?
\end{itemize}

In many physical systems there exists a single open CC consisting of an exponentially large number of configurations, leading to a unique stationary behavior in the thermodynamic limit. Examples include the contact process \cite{Harris1974}, directed percolation \cite{Kinzel1983,Hinrichsen2000}, the pair contact process \cite{Jensen1993}, some with experimental realizations \cite{Takeuchi2007,Takeuchi2009}. 
In this article we show that when there are many open communicating classes, each one thermodynamically large, they can compete, leading to stationary behavior that depends on the initial conditions. Furthermore, we find that the absorbing phase transitions occurring in these systems can exhibit memory-dependent scaling.
As an example, we demonstrate these ideas in a class of birth–death–diffusion (BDD) processes with the empty lattice as the unique absorbing configuration \cite{Collet2013,vanDoorn1991,Champagnat2016,Meleard2012}. The birth and death rates depend on the density. We employ a strong separation of time scales, birth–death events much slower than diffusion, to obtain an analytically tractable reduction to an effective one-dimensional random-walk dynamics; however, this separation is not required for the uniqueness or nonuniqueness of the stationary state. When the birth channel is active (nonzero birth rate), the nonabsorbing sector is effectively a single open CC, and the stationary state is unique. In the no-birth limit, the nonabsorbing sector splits into multiple open CCs (fixed-$N$ particle sectors), leading to nonunique stationary states that retain memory of the initial density profile. We also find that, in the no-birth limit, this memory persists at criticality, leading to initial-condition–dependent critical exponents. Later, we discuss the general criteria for nonunique stationary state and memory-dependence for general Markov processes with absorbing states.

\section{A birth–death–diffusion model}

We consider hard–core particles on a one–dimensional periodic lattice of size $L$ (sites $i=1,\dots,L$ with occupation variables $s_i\in\{0,1\}$) following a birth–death–diffusion dynamics,
\begin{equation}
  1 \xrightleftharpoons[B(\rho)]{D(\rho)} 0;\quad 10 \xrightleftharpoons[1]{1} 01.
  \label{eq:BD}
\end{equation}
Note that the on–site birth and death events occur with rates that depend on the instantaneous global density $\rho=N/L$ of the system, where $N=\sum_i s_i$. This, in a simple way, takes care of the interacting nature of the particle: addition and removal of a particle at a site depend on the occupancy of other particles in the neighborhood, which is a function of $\rho$. In addition, we focus on a class of rates,
\begin{equation}\label{eq:BDratesGeneral}
  D(\rho)= e^{-L\,\theta(\rho)}, \qquad B(\rho)= \rho\, e^{-L\,\vartheta(\rho)},
\end{equation}
so that $B(0)=0$. This condition ensures that particles cannot be added when the lattice is empty. Naturally, the empty lattice is an absorbing configuration and it is the only absorbing configuration of the system. We also assume a strong separation of time scales: birth–death events are exponentially slow in $L$ (typical waiting times $\sim e^{+L}$), in comparison to diffusion, which mixes particles on the lattice at unit rate. Such a separation of time scales allows us to reduce the BDD dynamics to an effective random-walk dynamics in particle-number ($N$) space and thereby helps us track the problem analytically.

For any finite $L$, following the dynamics in Eq.~\eqref{eq:BD}, the system will certainly reach the absorbing state (empty lattice) in the long-time limit. Thus the steady state is trivial: the probability of all configurations with $N\ne 0$ is zero and that of the empty lattice $N=0$ (denoted $\mathcal C_0$) is unity. For large $L$, however, the system can reside at $N\ne 0$ for a long period, leading to a quasi-stationary state (QS) with nonzero probability measure on the surviving configurations.

Formally, the quasi-stationary measure is defined by \cite{Darroch1965,vanDoorn1991,Meleard2012,Collet2013,Champagnat2016}
\begin{eqnarray}
\mathbb Q(\mathcal C) &=& \lim_{t\to\infty} \frac{\mathbb P(\mathcal C,t)}{\mathbb S(t)} \quad \forall\ \mathcal C \ne \mathcal C_0, \label{eq:QS}
\end{eqnarray}
where
\begin{equation}
\mathbb S(t) \equiv \sum_{\mathcal C\neq \mathcal C_0} \mathbb P(\mathcal C,t)
= 1 -  \mathbb P(\mathcal C_0,t), \label{eq:Surv}
\end{equation}
and the probability measure $\mathbb P(\mathcal C,t)$ obeys the master equation of the system for all $\mathcal C$. Clearly, Eqs.~\eqref{eq:QS} and \eqref{eq:Surv} ensure $\sum_{\mathcal C\neq \mathcal C_0}\mathbb Q(\mathcal C)=1$. What is not obvious is why, in general and in the long-time limit, $\mathbb Q(\mathcal C)$ is independent of time $t$ and of the initial condition. A proof of this, for a system with a single open communicating class, is given in Appendix~\ref{AppendixSec:UniquenessGeneral}. This proof also provides a clue as to when the QS state can become initial-condition dependent.

For the dynamics in Eq.~\eqref{eq:BD}, the quasi-stationary state can be obtained analytically from a mapping of the model to a random walk in particle-number space, i.e., on a one-dimensional lattice of $L+1$ sites labeled by $N=0,1,\dots,L$, with closed boundary at $N=L$ and an absorbing boundary at $N=0$:
\begin{equation}\label{eq:RW}
  N-1 \xleftarrow{\textstyle p_N}\; N \xrightarrow{\textstyle q_N}\; N+1
\end{equation}
where
\begin{equation}
  q_N = (L-N)\, B(\rho), \qquad p_N = N\, D(\rho), 
  \label{eq:BNDN}
\end{equation}
and the prefactors count, respectively, empty sites available for birth and occupied sites available for death. The walker moves to the left ($N\to N-1$) or right ($N\to N+1$) with rates $p_N$ and $q_N$, respectively, and stops if it hits $N=0$. The rates in Eq.~\eqref{eq:BNDN} ensure that birth is not possible at $N=L$ and that death is not possible at $N=0$. This mapping is possible because of the separation of time scales between diffusion and birth–death events.

Since the birth–death rates are much slower (of order $e^{-L}$) than diffusion (unit rate), the system has sufficient time to relax via diffusion before a change in particle number occurs. This allows us to assume that all configurations of the system with a fixed $N$ are visited (by diffusion) with equal probability,
\begin{equation}\label{eq:probConfigFixDensity}
  g_N(\mathcal C)= \binom{L}{N}^{-1}.
\end{equation}
where the binomial coefficient $\binom{L}{N}$ counts the configurations with $N$ particles.

This mapping allows us to obtain the QS distribution for the BDD process from $f_N$, the QS measure of the random walk defined in Eq.~\eqref{eq:BNDN}:
\begin{equation}\label{eq:factorizedProbMeasure}
  \mathbb Q(\mathcal C)= f_N\, g_N(\mathcal C).
\end{equation}
Then it is straight forward to calculate the  mean value of any observable $\cal O(\mathcal C)$  in the quasi-stationary state as  
\begin{equation}
 \langle  \mathcal O(\mathcal C) \rangle = \sum_{\mathcal C\ne \mathcal C_0} O(\mathcal C)  \mathbb Q(\mathcal C) 
\end{equation}

One of  our  goal here is to examine the possibility of an absorbing phase transition  controlled by the birth-death rates.  The  order parameter  of the  system  is the average  density in the QS state 
\begin{eqnarray}
  \varrho &\equiv& \frac{\langle N \rangle}{L}
  =\frac 1L \sum_{\mathcal C\ne \mathcal C_0} N(\mathcal C)  \mathbb Q(\mathcal C)\cr
  &=& \frac 1L\sum_{N=1}^{L} \binom{L}{N} N f_N g_N(\mathcal C)
  =\sum_{N=1}^{L} \frac{N}{L}\, f_N.\label{eq:averageDensityFirstDef}
\end{eqnarray}
Here in the last line we use Eqs.  \eqref{eq:probConfigFixDensity} and \eqref{eq:factorizedProbMeasure}; the factor $\binom{L}{N}$  counts the number of configurations  of the system  having $N$ particles. Clearly a state with  $\varrho=0$  is possible when the system  is in the absorbing configuration ${\mathcal C_0};$ in contrast to  $\varrho\ne0$ state  where the system  survives with   nonzero  probability in    the  active  configurations   ${\mathcal C \ne  \mathcal C_0}.$ 

To proceed further, i.e., to determine $f_N$, we need to specify the functional form of the birth–death rates, Eqs.~\eqref{eq:BDratesGeneral} and \eqref{eq:BNDN}, which we do in the next section.

\section{Absorbing phase transition}
\label{sec:uniqueAbsorbingTransition}

We start with a concrete example with the functional form of birth–death rates in \eqref{eq:BDratesGeneral} as
\begin{equation}
D(\rho) = e^{-L\rho(2d-\rho)}, 
\qquad 
B(\rho) = \rho\,e^{-L\rho(2/b-\rho)}, 
\label{eq:BDex}
\end{equation}
where real parameters $d$ and $b$. For $b>0$, both  the  birth and the death rates are nonzero for $N\neq 0$ and   all configurations   having $N\neq 0$  particles   can   be arrived starting from any other  $N\neq 0$ configurations; thus all  $N\neq 0$ sectors   together form a single open CC.
In  $b\to0^+$ limit, however,   birth rate  vanishes  creating  each  $N\neq 0$ sector a separate open CC. In the following  we will discuss the case $b>0$; $b=0$  case   will be  discussed in the next section (\autoref{sec:NoBirth}). 

The   birth-death dynamics   \eqref{eq:BDex}  leads  to  RW dynamics 

\begin{equation}
  N-1 \xleftarrow{\textstyle p_N}\; N \xrightarrow{\textstyle q_N}\; N+1
\end{equation} 
in $N$ space  with  hopping rates are given by  (from  Eq.~\eqref{eq:BNDN}),
\begin{equation}
  q_N = (L-N)B\left(\frac NL\right),\quad p_N = ND\left(\frac NL\right).
  \label{eq:BNDNex}
\end{equation}
Clearly, the site $N=0$ is an absorbing boundary since $q_0=0$. The evolution of the RW is governed by the master equation
\begin{equation}\label{eq:MasterEquation}
\frac{d}{dt}|P(t)\rangle = \mathbf M\,|P(t)\rangle,
\end{equation}   
where the $n$th component of $|P(t)\rangle$ is the probability of finding the RW at site $N=n$ at time $t$, and the $(L+1)\times(L+1)$ rate matrix $\mathbf M$ is
\begin{equation}
\mathbf{M}=\begin{pmatrix} 
	0 & p_1     & 0            & \dots & 0\\
	0 & -p_1-q_1& p_2          & \dots & 0\\
	0 & q_1     & -p_2-q_2     & \dots & 0\\
	0 & 0       & q_2          & \dots & 0 \\
	\vdots & \vdots & \vdots   & \ddots& \vdots\\
	0 & 0       & 0            & \dots & -p_L
\end{pmatrix}.
\end{equation}  

Label the eigenvalues of $\mathbf M$ in decreasing order of their real parts, $0=\lambda_0 \ge \Re(\lambda_1) \ge \Re(\lambda_2)\ge \dots$, and denote the left and right eigenvectors of $\lambda_n$ by $\langle \psi_n|$ and $|\psi_n\rangle$, respectively, 
  satisfying eigenvalue equation
\begin{equation}
  \langle \psi_n|\mathbf M =  \langle \psi_n|\lambda_n, \quad \mathbf M|\psi_n\rangle=\lambda_n |\psi_n\rangle.
\end{equation}

The steady state $|\psi_0\rangle = \begin{pmatrix} 1 & 0 & 0 & \dots \end{pmatrix}^T$ is trivial, in the limit $t\to\infty$ the system is certainly absorbed. Consider the initial condition localized at site $i\neq 0$: $|P(0)\rangle = |i\rangle$. Here $|i\rangle$ is the column basis vector with a 1 in the $i$‑th position and zeros elsewhere. The survival probability up to time $t$ is $\mathbb S_i(t) = \sum\limits_{k\neq 0}\langle k|e^{\mathbf M t}|i\rangle$, where $\langle k|$ is a row vector with a 1 in the $k$‑th position (zeros elsewhere). Conditional on survival, the probability of finding the walker at site $j\neq 0$ is
\begin{eqnarray}
  \mathbf{Q}_{ij}(t) &=& \frac{\,\langle j|e^{\mathbf M t}|i\rangle}{\mathbb S_i(t)},\\
  \displaystyle\mathbf{Q}_{ij}(t) &=& \frac{\sum\limits_{n} e^{\lambda_n t}\langle j|\psi_n\rangle\langle \psi_n|i\rangle}{\sum\limits_{k\neq 0}\sum\limits_{n} e^{\lambda_n t}\langle k|\psi_n\rangle\langle \psi_n|i\rangle},
\end{eqnarray}
  where we used the spectral decomposition $e^{\mathbf M t} = \sum\limits_{n} e^{\lambda_n t}|\psi_n\rangle\langle \psi_n|$. In the limit $t\to\infty$, since $\langle j\neq 0|\psi_0\rangle = 0$, the leading contribution comes from $n=1$, yielding
\begin{eqnarray}
  \lim_{t\to\infty} \mathbf{Q}_{ij}(t) &=&
\frac{\langle j|\psi_1\rangle \langle \psi_1|i\rangle e^{-\lambda_1t}}{ \sum\limits_{k\ne0} \langle k|\psi_1\rangle \langle \psi_1|i\rangle e^{-\lambda_1t}},\\
  \lim_{t\to\infty} \mathbf{Q}_{ij}(t) &=&
\frac{\langle j|\psi_1\rangle}{ \sum\limits_{k\ne0} \langle k|\psi_1\rangle }\equiv f_j,\label{Qij}
\end{eqnarray} 
where $f_j$ denotes the quasi-stationary (QS) weight at state $j\neq 0$. It is given (up to normalization) by the components of the right eigenvector $|\psi_1\rangle$ associated with the second-largest eigenvalue of the Markov generator $\mathbf M$, namely $|\psi_1\rangle = \begin{pmatrix} f_0 & f_1 & f_2 & \dots \end{pmatrix}^{\!T}$, with the normalization $\sum_{j\neq 0} f_j = 1$. Note that, the QS state is unique in the sense that it does not depend on the initial condition (here, the starting density $\rho=i/L$). The eigenvalue equation $\mathbf M |\psi_1\rangle = \lambda_1 |\psi_1\rangle$ yields the recursion
\begin{equation}
p_{n+1}f_{n+1} = \big(p_{n}+q_{n}+\lambda_1\big)f_{n} - q_{n-1}f_{n-1},
\label{eq:recur}
\end{equation} 
with boundary conditions $q_0=0$, $p_0=0$, and $q_L=0$. When $bd\le 1$ and $L$ is large, the eigenspectrum satisfies $\{\lambda_n\}\approx \{-p_i-q_i\}$, i.e., the diagonal entries of $\mathbf M$. With the largest eigenvalue $\lambda_0=0=-p_0-q_0$, the second-largest eigenvalue is $\lambda_1\approx -\min_{i\neq 0}(p_{i}+q_{i})$. When $bd>1$, one finds $\lambda_1\ll -\min_{i\neq 0}(p_{i}+q_{i})$. To determine the $f_i$, we solve the recursion in Eq.~\eqref{eq:recur} separately for the three cases: $bd=1$, $bd<1$, and $bd>1$.
\begin{enumerate}
  \item For $d=1/b$, $q_n/p_n = 1 - n/L$ is ${\mathcal O}(1)$ and hence $q_n \simeq p_n$. Equation~(\ref{eq:recur}) then gives the solution
\begin{equation}
f_n = \frac{n\,\Theta(d-\rho)}{p_n} \;\simeq\; \Theta(d-\rho)\, e^{L \rho(2d -\rho)}.
\end{equation}
We obtain the asymptotic form by taking the large-$L$ limit: for any positive polynomial $h(\rho)$, 
$h(\rho)e^{L\theta(\rho)} = e^{L\theta(\rho)+\ln h(\rho)} \approx e^{L\theta(\rho)}$ when $L$ is large. We use this Laplace (saddle-point) approximation repeatedly throughout this article. The Heaviside step function $\Theta(d-\rho)$ ensured that the iteration of \eqref{eq:recur} ends when $p_n+q_n+\lambda_1=0$, i.e., when $n=\arg\min\limits_i\,(p_i+q_i)=Ld$.

  \item For $bd<1$, $q_i \ll p_i$ and Eq.~(\ref{eq:recur}) reduces to $p_n f_n = p_{n-1} f_{n-1}$, yielding
\begin{equation}
f_n = \frac{\Theta(d-\rho)}{p_n}= \Theta(d-\rho)\, e^{L \rho(2d -\rho)}.
\end{equation}
  \item For $bd>1$, $q_i \gg p_i$, and $p_i+q_i+\lambda_1$ in Eq.~(\ref{eq:recur}) can be replaced by $q_i$. This leads to
\begin{equation}
f_n \simeq \frac{1}{p_n}\prod_{k<n}\frac{q_k}{p_k}
\simeq e^{L\rho(2d-\rho)+L^2(d-1/b)\rho^2}.
\end{equation}
\end{enumerate}

\begin{figure}[t]
\includegraphics[width=\columnwidth]{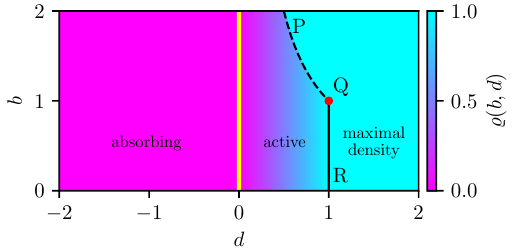} 
\caption{
  \textbf{Absorbing phase diagram in the birth–death–diffusion process.}
  Quasi–stationary steady states are classified by the density $\varrho$: absorbing ($\varrho=0$), active ($0<\varrho<1$), and maximal density ($\varrho=1$). The APT occurs at $d=0$. Along the PQR line separating active and maximal-density phases, the complementary order parameter $\psi\equiv 1-\varrho$ exhibits an ordinary transition: it is continuous (second order) on QR and discontinuous (first order) on PQ, meeting at a tricritical point $Q$ where $b=d=1$.
}\label{fig:phases}
\end{figure}

In summary, in the thermodynamic limit, the QS measure of the RW dynamics (\ref{eq:RW}) is
\begin{equation}\label{eq:QSmeasureWithBirth}
f(\rho)=
\begin{cases}
\Theta(d-\rho)\,e^{L \rho(2d -\rho)}, & bd \le 1,\\[8pt]
 e^{L^2 \rho^2 (d-1/b)}, & bd >  1,
\end{cases}
\end{equation}
The average steady-state density is
\begin{equation}
\varrho = \langle\rho\rangle =\frac{1}{\mathcal N} \int_0^1 \rho\, f(\rho)\, d\rho.
 \end{equation}
where $\mathcal N = \int\limits_0^1 f(\rho)\, d\rho $.
Note that  the  above equation is  the continuum form of Eq. \eqref{eq:averageDensityFirstDef} discussed in the previous section. 
Now the orderparamter $\varrho$ of the ansorbing  transition can be evaluated straightforwardly using the steepest-descent method \cite{Wong1989}, yielding 
\begin{equation}\label{eq:rho_s}
\varrho =
\begin{cases*}
0, & $d<0$,\\
d, & $0\le d<1$ and $bd \le 1$.\\
1, & otherwise.
\end{cases*}
\end{equation}
Clearly $\varrho$ picks up nonzero value when $d>0.$  Thus the BDD model exhibits a continuous absorbing transition along the line $d=0$.
Fig. \ref{fig:phases} displays  the  density plot of $\varrho$ in   
 $d$--$b$ plane, with  the critical line  $d=0$ (thick solid line).  

We characterize the critical behavior with respect to the control parameter \(d\) near its critical value \(d_c = 0\). We denote the distance to criticality by
\begin{equation}
  \Delta \equiv d - d_c \quad (d_c=0),
\end{equation}
In the limit \(\Delta \to 0^+\), the order parameter and the susceptibility (defined as \(\chi \equiv L\,\mathrm{Var}(\rho)\)) obey the finite-size scaling forms
\begin{align}
\varrho \equiv \langle \rho \rangle
&= L^{-\beta/\nu}\, \mathcal{F}_1(\Delta L^{1/\nu}),
\label{eq:rho-scaling}\\
\chi \equiv L\big(\langle\rho^2\rangle-\langle\rho\rangle^2\big)
&= L^{\gamma/\nu}\, \mathcal{F}_2(\Delta L^{1/\nu}).
\label{eq:chi-scaling}
\end{align}
For the BDD model, the exact expressions yield
\begin{align}
\varrho &= L^{-1/2}\, \mathcal{F}_1(\Delta L^{1/2}),
\label{eq:rho-scaling-BDD}\\
\chi &= \mathcal{F}_2(\Delta L^{1/2}),
\label{eq:chi-scaling-BDD}
\end{align}
with scaling functions
\begin{align}
\mathcal{F}_1(x)
&= x + \frac{e^{-x^2} - 1}{\sqrt{\pi}\,\operatorname{erf}(x)},\\
\mathcal{F}_2(x)
&= \frac{1}{2}
- \frac{e^{-x^2}\!\left(-2 + 2\cosh(x^2) + x \sqrt{\pi}\,\operatorname{erf}(x)\right)}
{\pi\, \operatorname{erf}(x)^2}.
\end{align}
These functions are regular at \(x=0\) and have the limits
\begin{eqnarray}
\mathcal{F}_1(0)&\,=\,\dfrac12, \quad \mathcal{F}_2(0)&\,=\,0;
\\
\mathcal{F}_1(x)&\,\to\,x, \quad \mathcal{F}_2(x)&\,\to\,\frac{1}{2}-\frac{1}{\pi}, \;\; (x\to\infty).
\end{eqnarray}
Comparing Eqs.~\eqref{eq:rho-scaling}–\eqref{eq:chi-scaling} with Eqs.~\eqref{eq:rho-scaling-BDD}–\eqref{eq:chi-scaling-BDD}, we identify the critical exponents as
\begin{equation}
  \beta = 1,\qquad \gamma = 0,\qquad \nu = 2.
  \label{eq:uniq_exp}
\end{equation}
They satisfy the hyper scaling relation \(\gamma = \nu - 2\beta\) \cite{Stanley1972}.

In addition to   the absorbing phase transition that  occurs  at $d=0,$   another phase transition occurs   along  line PQR  shown  in Fig.  \ref{fig:phases}:  the  transition line  PQR   separates  the active phase  ($0<\varrho<1$) 
from the maximally active region $\varrho=1.$   A natural order parameter of  this transition is   $\psi\equiv 1-\varrho$ as the order parameter which picks up nonzero value  only in the $0<\varrho<1$ region.  The equation of the  line PQR is given by 
\begin{equation}
\overline d_c(b) =
\begin{cases}
1, & 0<b<1,\\
b^{-1}, & b\ge 1.
\end{cases}
\end{equation}
The transition along the line PQ is discontinuous because $\psi$ jumps from $0$ to $1-1/b$ at $d=\overline d_c$. And along the line QR, the transition is continuous. Thus the point $Q$   with coordinate $(d,b)=(1,1)$ is a tricritical point.

\begin{figure}[t]
\includegraphics[width=\columnwidth]{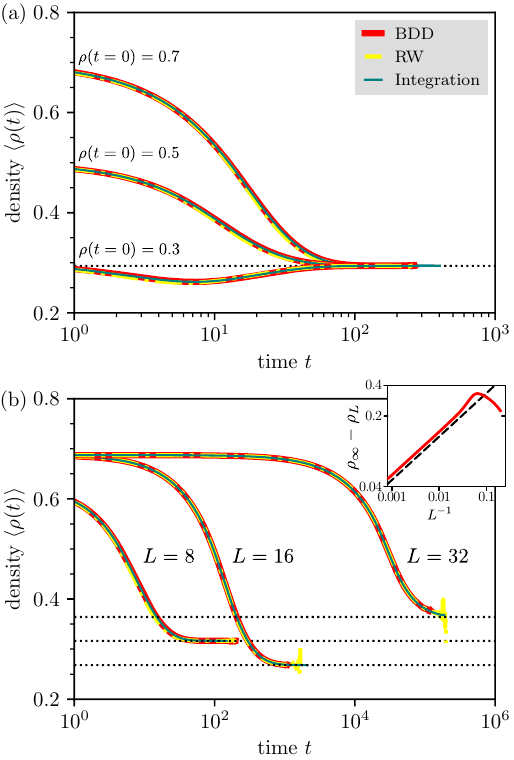}
\caption{
  \textbf{Reduction of a birth–death–diffusion process to an effective random-walk description.}
  Time evolution of the density $\langle\rho(t)\rangle$ for the rates in Eq.~(\ref{eq:BDex}) with $b=1.4$ and $d=0.6$. 
  \textbf{(a)} Decay for $L=10$ with initial densities $\rho(0)=0.3,\,0.5,\,0.7$. 
  \textbf{(b)} Decay for $L=8,16,32$ initialized with $N(0)=\lfloor 0.7L\rfloor$. 
  In each panel, three overlapping curves show Monte Carlo simulation of the full BDD model (solid red), simulation of the effective RW (dashed yellow), and numerical integration of the RW master equation (solid teal), demonstrating the validity of the reduction. Horizontal dotted lines mark $\varrho=\langle\rho(t\to\infty)\rangle$ obtained using the quasi–stationary simulation method of Ref.~\cite{Oliveira2005}. Inset of (B): finite–size approach of $\varrho_L$ to the asymptotic value $\varrho_{\infty}=0.6$ in agreement with Eq.~(\ref{eq:rho_s}); the deviation follows $\varrho_{\infty}-\varrho_L\sim L^{-0.428}$ (dashed guide line).
}\label{fig:rho_L}
\end{figure}

The  results we  derive here is based on the mapping of the BDD model to an effective random walk   in $N$-space.  Given  an exponential   separation of time scales  between  diffusion and  birth-death, the validity of  this approximation is  intuitively clear.  As  an explicit check we  calculate the  evolution of mean density $\langle\rho(t)\rangle$ of the BDD  model Eq. \eqref{eq:BD}, the Random walk model  Eq. \eqref{eq:RW}  and the same  calculated directly from  integrating the Master equation \eqref{eq:MasterEquation},
\begin{equation}
  \langle\rho(t)\rangle = \sum_{\mathcal C\notin A} \rho(\mathcal C)\, \frac{\mathbb P(\mathcal C,t)}{\mathbb S(t)}.
\end{equation}
This is shown in Fig.~\ref{fig:rho_L} for $b=1.4$ and $d=0.6.$

Figure 3(a) shows the time evolution of the average density $\langle \rho(t) \rangle$ for $L=10$, starting from different initial conditions. In all cases, the dynamics relaxes toward the same stationary value, $\langle \rho \rangle \simeq 0.294$, indicating that the long-time behavior is independent of the initial state. The dashed horizontal line in Fig.~3(a) denotes this stationary value obtained using the quasistationary (QS) simulation method of Ref.~\cite{Oliveira2005}.

In the QS method, a running histogram $H(\rho)$ of the particle density is maintained by recording the values of $\rho$ sampled during the active portions of the trajectory, i.e., whenever the system satisfies $\rho>0$. If the BDD process (1) becomes absorbed and reaches the empty lattice ($\rho=0$), the system is immediately reinitialized in a nonabsorbing configuration. This is done by first drawing a density value $\rho_0$ from the histogram $H(\rho)$, which represents densities previously visited during the active evolution. A configuration with this density is then generated by placing $\rho_0 L$ particles randomly and uniformly on the lattice. The stochastic dynamics is subsequently resumed, and the histogram $H(\rho)$ continues to be updated throughout the simulation.

As the simulation proceeds, $H(\rho)$ progressively approaches the QS distribution $f(\rho)$,i.e., $H(\rho)\to f(\rho).$  In this way, the QS method prevents the system from becoming permanently trapped in the absorbing state while preserving the correct statistical properties of the active phase, thereby enabling efficient sampling of the QS regime~\cite{Oliveira2005}.

Note that, the saturation value of $\langle \rho\rangle=0.294$ is different from the expected value $\varrho = d = 0.6.$ In fact, for different system sizes $L$, we observe that $\langle \rho\rangle$ saturates to different values, see Fig.~\ref{fig:rho_L}(b). This is  a finite size effect  which can be corrected  as we approach larger  $L.$ Simulating a system   with rates  ${\cal O}(e^{-L})$  is  computationally expensive for large $L$. To verify that  the system  indeed approach  the QS state having density  $\varrho_\infty = d = 0.6$  in  $L\to \infty$ limit, using the quasi–stationary simulation method of Ref.~\cite{Oliveira2005},  we  calculate   $\varrho_L$ for  different $L$ and plot  
$\varrho_\infty -\varrho_L$ vs $L$   in log-scale  (see  the inset in Fig. \ref{fig:rho_L} (b)). 
This figure clearly indicates  that $\varrho_L$  indeed  approaches $\varrho_\infty$  in   the thermodynamic limit.

Note that the quasi stationary states we studied here for $b>0$ is unique  and  the  absorbing phase transition is characterized by a unique set of critical exponents, Eq. \eqref{eq:uniq_exp}. This  is due to the fact  that  all  the non-absorbing configurations form  a single dynamically connected set,  namely {\it an open communicating class}. The generator associated with  this set is irreducible and  it admits a unique QS measure, independent of the initial condition. The proof of this  statement is based on   the Perron–Frobenius theorem, discussed   in detail in  Appendix \ref{AppendixSec:UniquenessGeneral}. By contrast, in the limit $b\to 0^{+}$, the  births events are switched off and then the  set of  non-absorbing  configurations   splits into multiple open communicating classes, each corresponding to  the set of configurations having  exactly $N$-particles. The restricted generator is then reducible, the Perron–Frobenius theorem no longer  defend a  unique positive eigenvector for the largest eigenvalue  and   corresponding QS state need not be unique --it can retain  the  memory of  the initial condition. This  special case $b=0$ is analyzed in detail in the next section (\autoref{sec:NoBirth}).

\section{Non-unique quasi–stationary state in the absence of birth}
\label{sec:NoBirth}

The line $b=0$ corresponds to vanishing birth rate. In this case the master equation of the BDD process \eqref{eq:BD} can be solved exactly on the full configuration space ($2^L$ configurations), which is composed of $L+1$  sectors, each corresponding to a fixed particle number $N=0,1,\dots,L.$   A sector   with $N$ particles  has  $\binom{L}{N}$ configurations in total.
 The master equation  for $b=0$ reads as 
\begin{equation}
  \frac{d}{dt}\,|P\rangle \;=\; \mathbf{M}\,|P\rangle,\qquad
|P\rangle \equiv \big(|\phi_0\rangle,|\phi_1\rangle,\dots,|\phi_L\rangle\big)^{\!T},
\end{equation}
where  $|\phi_N\rangle$  is a  column vector of dimension $\binom{L}{N},$  which makes $|P\rangle$  a   column vector of dimension $2^L.$  Since the birth rate $B=0,$  the rate matrix $\mathbf M$ of Markov process takes a block upper–triangular form
\begin{equation}
\mathbf{M} \;=\;
\begin{pmatrix}
\mathbf{M}_0      & \mathbf{P}_1 & \mathbf{0}          & \mathbf{0}          & \dots & \mathbf{0}\\
\mathbf{0}        & \mathbf{M}_1     &  \mathbf{P}_2    & \mathbf{0}          & \dots & \mathbf{0}\\
\mathbf{0}        & \mathbf{0}       & \mathbf{M}_2        & \mathbf{P}_3    & \dots & \mathbf{0}\\
\vdots            & \vdots           & \vdots              & \vdots              & \ddots& \vdots\\
\mathbf{0}        & \mathbf{0}       & \mathbf{0}          & \mathbf{0}          & \dots & \mathbf{M}_L
\end{pmatrix}.
\end{equation}
where
\begin{equation}
    \mathbf M_n = \mathbf C_n - nD_n\,\mathbf I_n,\quad
    \mathbf P_n = D_n\,\mathbf T_n,\quad 
    D_n\equiv D(n/L).
\end{equation}
The diagonal blocks $\mathbf M_n$ are $\binom{L}{n}\times \binom{L}{n}$ and encode conservative diffusion $\mathbf C_n$ within the $n$–particle sector together with the total loss $-nD_n$ due to particle death. The off–diagonal blocks $\mathbf P_n$ implement the number change $n\to n-1$; here $\mathbf T_n$ is the $\binom{L}{n-1}\times \binom{L}{n}$ deletion matrix that maps $n$–particle configurations to $(n-1)$–particle configurations by removing one particle. Since birth is absent, there are no transitions from lower to higher sectors, and the empty lattice ($n=0$) is absorbing.

Each column of $\mathbf T_n$ contains exactly $n$ ones (the $n$ possible deletions), which implies
\begin{equation}\label{eq:Tk-ones-correct}
\langle 1_{n-1}|\,\mathbf T_n \;=\; n\,\langle 1_{n}|,
\end{equation}
where $\langle 1_k|\equiv(1,1,\dots,1)$ is the all-ones row vector of length $\binom{L}{k}$. Probability conservation diffusion dynamics $\mathbf C_n$ within each fixed-$n$ sector implies
\begin{equation}
\langle 1_n|\,\mathbf C_n \;=\; 0.
\end{equation}
The uniform stationary state in the $n$–particle sector is
\begin{equation}
  |\overline\phi_n\rangle = g_n\,|1_n\rangle,\qquad
g_n=\binom{L}{n}^{-1},
\end{equation}
with $|1_n\rangle=(1,1,\dots,1)^{T}$ (length $\binom{L}{n}$ column vector) and
\begin{equation}
  \mathbf C_n\,|\overline\phi_n\rangle = 0.
\end{equation}

Since $\mathbf M$ is block upper–triangular, its spectrum is the union of the spectra of the diagonal blocks,
\begin{equation}
\Lambda(\mathbf M) \;=\; \bigcup_{n=0}^{L}\Lambda(\mathbf M_n),
\quad
\lambda_{n\nu} \;=\; \Lambda(\mathbf C_n)_\nu \;-\; nD_n,
\end{equation}
where $\Lambda(\mathbf C_n)_\nu$ are ordered by decreasing real part, $\nu=0,1,\dots,\binom{L}{n}$. We are primarily interested in the leading mode in each sector, $\lambda_n\equiv\lambda_{n0}=-\,nD_n$.

Write the left and right eigenvectors of $\lambda_n$ as concatenations over sectors,
\begin{equation}
\langle\Psi_n|=\big(\langle\psi_{n,0}|,\langle\psi_{n,1}|,\dots\big),\quad|\Psi_n\rangle=\begin{pmatrix}|\psi_{n,0}\rangle\\
    |\psi_{n,1}\rangle\\
    \vdots
  \end{pmatrix}.
\end{equation}
The right–eigenvalue equations $\mathbf M |\Psi_n\rangle = \lambda_n\,|\Psi_n\rangle$ lead to the recursion
\begin{equation}
\mathbf P_{i+1}|\psi_{n,i+1}\rangle + \mathbf M_i|\psi_{n,i}\rangle
= \lambda_n\, |\psi_{n,i}\rangle
\qquad (i\ge 0),
\end{equation}
which can be solved recursively. We find right–eigenvectors corresponding to $\lambda_n$ by substituting $|\psi_{n,n}\rangle = |\bar\phi_n\rangle$. This leads to the compact forms
\begin{align}
|\psi_{n,k}\rangle &=
\begin{cases}
\displaystyle \prod_{i=1}^{n-1}\frac{iD_i}{\,nD_n - iD_i\,}, & k=0,\\[8pt]
\displaystyle \prod_{i=k+1}^{n}\,(\lambda_n\mathbf I_{\,i-1}-\mathbf M_{\,i-1})^{-1}\,\mathbf P_i\,|\overline\phi_n\rangle, & 0<k\le n,\\[6pt]
\mathbf 0, & k>n,
\end{cases}\label{eq:psi-right-correct}
\end{align}
where $\mathbf 0$ denotes a null vector of the appropriate dimension. The $k=0$ component in (\ref{eq:psi-right-correct}) follows by repeated use of (\ref{eq:Tk-ones-correct}) and
\begin{equation}
  \langle 1_i|\big(\mathbf M_i-\lambda_n\mathbf I_i\big)^{-1}
  \;=\; \frac{\langle 1_i|}{\,nD_n - iD_i\,}.
\end{equation}

Similarly, the left–eigenvalue equations $\langle \Psi_n|\, \mathbf M  = \langle \Psi_n|\,\lambda_n$ give
\begin{equation}
  \langle \psi_{n,i-1}|\mathbf P_{i} + \langle \psi_{n,i}|\mathbf M_i  = \lambda_n\, \langle \psi_{n,i}|\qquad (i\ge 0),
\end{equation}
and with $\langle\psi_{n,n}| = \langle\overline\phi_n|$ one finds
\begin{align}
  \langle\psi_{n,k}| &=
\begin{cases}
\mathbf 0, & k<n,\\
\displaystyle \langle 1_k| \prod_{i=n+1}^{k}\frac{iD_i}{\,iD_i + \lambda_n\,}, & k\ge n.
\end{cases}\label{eq:psi-left-correct}
\end{align}

Using (\ref{eq:psi-left-correct}), we compute the overlap of $\langle\Psi_{\bar k}|$ with the initial condition $|m\rangle$ (a uniform distribution over all configurations with $m$ particles):
\begin{equation}\label{eq:overlap-correct}
  \langle \Psi_{\bar k} | m \rangle \;=\; \begin{cases}
0, & m< \bar k,\\[4pt]
\displaystyle \prod\limits_{i=\bar k+1}^{m}\frac{iD_i}{iD_i+\lambda_{\bar k}}, & m \geq \bar k,
\end{cases}
\end{equation}
where $| m \rangle =g_m\big(\mathbf 0,\dots, | 1_m\rangle,\dots,\mathbf 0\big)^T$. Thus the overlap $\langle \Psi_{\bar k} | m \rangle = 0$ vanishes for $m< \bar k$, and the overlap $\langle \Psi_{\bar k} | m \rangle > 0$ for $m\ge \bar k$, i.e., the sector index $\bar k$ does not exceed the initial particle number $m$.

Let $\mathbf{Q}_{mn}(t)$ be the conditional probability that, starting from a configuration with $m$ particles at $t=0$, the system has not been absorbed by time $t$ and has $n$ particles at time $t$:
\begin{eqnarray}
  \mathbf{Q}_{mn}(t) &=& \frac{\langle n|e^{\mathbf Mt}|m\rangle}{1-\langle 0|e^{\mathbf Mt}|m\rangle}\\
  &=& \frac{\sum\limits_{k\ge 1} e^{-kD_k t}\,\langle n|\Psi_{k}\rangle\,\langle\Psi_{k}|m\rangle}{-\sum\limits_{k\ge 1} e^{-kD_k t}\,\langle 0|\Psi_{k}\rangle\,\langle\Psi_{k}|m\rangle},
\end{eqnarray}
where we define $\langle n| = \big(\mathbf 0,\dots, \langle 1_n|,\dots,\mathbf 0\big)$. Only modes with $k\le m$ contribute because $\langle\Psi_k|m\rangle=0$ for $k>m$. In the long-time limit, the dominant contribution comes from the mode with the slowest decay among those that overlap the initial sector,
\begin{equation}\label{eq:initial-condition-kbar}
  \bar k \;=\; \arg\min_{1\le k\le m} \big(kD_k\big),
\end{equation}
so long as the minimizer is unique. In limit $t\to\infty$, we find
\begin{eqnarray}
  \lim_{t\to\infty}\mathbf{Q}_{mn} &=& \frac{\langle n|\Psi_{\bar k} \rangle}{-\langle 0|\Psi_{\bar k} \rangle}\\
  &=& \begin{cases}
\displaystyle \frac{\bar kD_{\bar k}}{nD_n}\prod\limits_{k<n}\left(1-\frac{\bar kD_{\bar k}}{kD_k}\right), & n\le \bar k(m),\\[10pt]
0, & n>\bar k(m).
\end{cases}\label{eq:Qmn-product-correct}
\end{eqnarray}
The initial–condition dependence appears via $\bar k(m)$ from Eq.~\eqref{eq:initial-condition-kbar}.

To obtain the continuum form of $\mathbf{Q}_{mn}$, write $m=\rho_{\mathrm{in}} L$, $n=\rho L$, $\bar k=\rho^* L$, and $D_k=D(\rho_k)$ with $\rho_k=k/L$. The key product in (\ref{eq:Qmn-product-correct}) is
\begin{align}
\prod_{k<\bar k}\left(1-\frac{\bar kD_{\bar k}}{kD_k}\right)
&\simeq& \prod_{k<n}\Big(1- \exp\big\{L[\theta(\rho_k)-\theta(\rho^*)]\big\}\Big) \\
& = &\exp\!\left[\sum_{k<n}\ln\!\Big(1- e^{L(\theta(\rho_k)-\theta(\rho^*))}\Big)\right]. \nonumber
\end{align}
The expression above is not a Riemann sum. Consider instead \eqref{eq:riemann-sum}:
\begin{widetext}
  \begin{equation}\label{eq:riemann-sum}
\frac1L\sum_{k<n}\ln\!\Big(1- e^{L(\theta(\rho_k)-\theta(\rho^*))}\Big)
\;
\longrightarrow
\;\int_0^{\rho}\!\mathrm d\rho'\,\ln\!\Big(1- e^{L(\theta(\rho')-\theta(\rho^*))}\Big)
  \end{equation}
\end{widetext}
Evaluating \eqref{eq:riemann-sum} by Laplace’s (saddle-point) method around the minimizer of $\theta(\rho)-\theta(\rho^*)$, we obtain the asymptotic kernel
\begin{equation}
  \mathbf{Q}_{mn} \to Q(\rho_{\mathrm{in}},\rho)
  = \frac{1}{\mathcal N_{\mathrm{in}}}\exp\!\left(L\big[\theta(\rho)-\theta(\rho^*)\big]\right)\Theta(\rho^*-\rho),
\end{equation}
where \(\mathcal N_{\mathrm{in}}\) is the normalization ensuring \(\int_0^1 \mathrm d\rho\, Q(\rho_{\mathrm{in}},\rho)=1\), and \(\rho^*=\bar k/L\) (the slowest eigenvalue mode selected by the initial density \(\rho_{\mathrm{in}}\)) follows from Eq.~\eqref{eq:initial-condition-kbar} is
\begin{equation}\label{eq:65}
\rho^*=
\begin{cases}
\rho_{\mathrm{in}}, & \rho_{\mathrm{in}}\le d,\\
d, & \rho_{\mathrm{in}}>d\ge 0,\\
0, & d<0.
\end{cases}
\end{equation}

For an arbitrary initial‑density distribution \(\pi(\rho_{\mathrm{in}})\) with \(\rho_{\mathrm{in}}>0\), the ensemble distribution over \(\rho\) is
\begin{equation}
\mathbb{Q}(\rho)=\int_{0^+}^{1}\!\mathrm d\rho_{\mathrm{in}}\, \pi(\rho_{\mathrm{in}})\, Q(\rho_{\mathrm{in}},\rho).
\end{equation}
Here we define \(\mathbb{Q}\) as the mixture of QS limits with differnet initial condition: first take the \(\rho_{\mathrm{in}}\)‑conditioned limit to obtain \(Q(\rho_{\mathrm{in}},\rho)\), then average over \(\rho_{\mathrm{in}}\) with weight \(\pi(\rho_{\mathrm{in}})\). This is distinct from pooling all long‑time surviving trajectories initialized with \(\pi(\rho_{\mathrm{in}})\), which implicitly reweights by survival and can bias toward the slowest sector.

All moments of the QS density distribution are
\begin{align}
  \langle \rho^k\rangle
  &= \int_0^1 \mathrm d\rho\, \mathbb{Q}(\rho)\, \rho^k
  \;\approx\; \int_0^1 \mathrm d\rho_{\mathrm{in}}\, \pi(\rho_{\mathrm{in}})\, \big(\rho^*\big)^k,
\end{align}
where we again used Laplace’s method \cite{Wong1989}. For the family of initial distributions
\begin{equation}
\pi(\rho_{\mathrm{in}})=(1+a)\,\rho_{\mathrm{in}}^a,\qquad a>-1,
\label{eq:init_a}
\end{equation}
the moments admit a closed form:
\begin{equation}
  \langle \rho^k\rangle \;=\; d^k \;-\; \frac{k}{\,k+1+a\,}\, d^{\,k+1+a},\quad k\ge 0.
\end{equation}
Thus the initial condition (via the exponent \(a\)) persists in the moments of the density and controls the subleading scaling near criticality at the critical point \(d_c=0\). This initial‑condition is imprinted on scalling exponents. For \(\Delta\equiv d-d_c\to 0^+\),
\begin{equation}
\varrho=\langle\rho\rangle \sim \Delta^{\beta},\qquad \beta=1, \label{eq:beta}
\end{equation}
while the fluctuation scales as
\begin{equation}
\langle\rho^2\rangle-\langle\rho\rangle^2=
\frac{d^{a+3}}{a+2} \left(\frac{2}{a+3}-\frac{d^{a+1}}{a+2}\right)
\sim \Delta^{-\gamma};~\gamma=-(a+3), \label{eq:gamma}
\end{equation}
yielding a nonunique susceptibility exponent \(\gamma\) controlled by the preparation exponent \(a\). We expect critical exponent $\nu$ to follow relation $\nu = \gamma + 2\beta$.
So the correlation‑length exponent inherits the same initial‑condition dependence. Continuous variation of exponents is known in systems with marginal parameters  
are  known   theoretically \cite{Kosterlitz1974, Baxter1971, Pearce1987, Indranil2023} and  some indication has been  observed   in  experiments \cite{Suzuki1992, Khan2017}. An explicit dependence  of  critical exponents on the initial  condition  is a distinctive feature of this BDD class of models at the  absorbing phase transition.

\begin{figure}[t]
\includegraphics[width=\columnwidth]{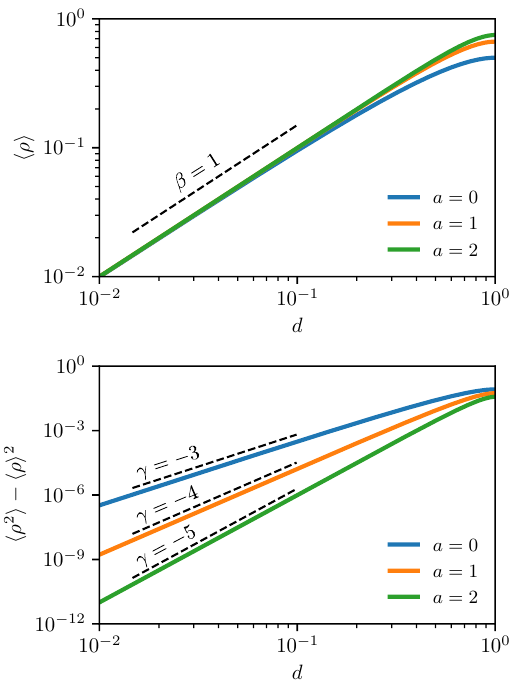}
\caption{ 
  \textbf{Initial-condition–dependent scaling at the absorbing transition.}
   A plot of $\langle\rho\rangle$ and its fluctuation $\langle\rho^2\rangle-\langle\rho\rangle^2$ versus $d,$  calculated in Eqs. \eqref{eq:beta} and \eqref{eq:gamma} for initial distributions $\pi(\rho)=(1+a)\rho^a$ are shown   for  $a=0,1,2$. For $d_c\approx 0$, $\langle\rho\rangle\sim d^{\beta}$ with $\beta=1$, and $\langle\rho^2\rangle-\langle\rho\rangle^2\sim d^{-\gamma}$ with $\gamma=-(a+3)$, showing initial-condition dependence of the susceptibility exponent (dashed line).
  }
\label{fig:rhos}
\end{figure}

The  results we obtain   in  this section (also in \autoref{sec:uniqueAbsorbingTransition})  are robust in the sense that  they    
are insensitive  to the choice of conservative dynamics.   For example, introducing   an attractive   nearest neighbour interaction like   Kawasaki spin‑exchange (Ising) to replace the   diffusion dynamics Eq. \eqref{eq:BD} :
\begin{equation}
0100\xrightleftharpoons[1]{1}0010,
0101\xrightleftharpoons[\alpha]{1}0011,
1100\xrightleftharpoons[1]{\alpha}1010,
1101\xrightleftharpoons[1]{1}1011,
\end{equation}
with \(\alpha=e^{-2J}\),  does  not change the qualitative behaviour of the model. In fact, both the phase diagram
(Fig. \ref{fig:phases})  and the   stationary state  (Eqs. \eqref{eq:rho_s} and \eqref{eq:65}) remains  the same. This is because the topological structure of communicating classes is not changed.

\section{Criteria for nonunique quasi-stationary states}
\label{sec:CriteriaNonuniqueQS}

In a Markov chain, two configurations belong to the same communicating class (CC) if each is accessible from the other. A CC is closed if it cannot be left; otherwise it is open. If the nonabsorbing sector \(S^+=\Omega\setminus A\) is irreducible (i.e., a single open CC), then the generator restricted to \(S^+\) satisfies the Perron–Frobenius (PF) theorem and the quasi–stationary (QS) distribution on \(S^+\) is unique and strictly positive \cite{Keizer1972}. This situation occurs in many models with absorbing states, e.g., the contact process \cite{Harris1974}, directed percolation \cite{Kinzel1983,Hinrichsen2000}, and the pair contact process \cite{Jensen1993} (see Appendix~\ref{AppendixSec:UniquenessGeneral}).

By contrast, if \(S^+\) contains multiple open CCs, the subgenerator on \(S^+\) is reducible and the PF theorem does not apply to the full nonabsorbing block. 
In this case the QS state can be nonunique, and the QS distribution \(\mathbb{Q}\) can depend on the initial condition: different initial conditions can select different slowest overlapping modes (cf. Eq.~\eqref{eq:overlap-correct}) and hence different long–lived weights.

To make this precise, let \(\mathcal S_n\) denote the fixed–\(n\) open CC (the set of all configurations with \(n\) particles), and define the total (effective) outflux rate from \(\mathcal S_n\) as $\Omega_n$ which depends on the number of allowed transitions \(n\to i<n\) and their rates \(\omega_{n\to i}\). For the BDD dynamics in Eq.~\eqref{eq:BD} with \(B=0\), one finds \(\Omega_n = n\,D_n\) because each \(n\)–particle configuration admits \(n\) path to sector $\mathcal S_{n-1}$, each at death rate \(D_n\). For the slowest overlapping mode with initial state \(\bar k(m)\), the QS measure \eqref{eq:Qmn-product-correct} takes the generic form
\begin{equation}\label{eq:Qmn-general}
  \lim_{t\to\infty}\mathbf{Q}_{mn} \;=\; \begin{cases} \displaystyle
\frac{\Omega_{\bar k}}{\Omega_n}\prod_{k<n}\left(1-\frac{\Omega_{\bar k}}{\Omega_k}\right),
& n\le \bar k(m),\\[10pt]
0, & n>\bar k(m),
\end{cases}
\end{equation}
where \(\bar k(m)\) minimizes \(\arg\min\limits_{1\le k\le m}\Omega_k\) among modes that overlap the initial sector \(m\). In the thermodynamic limit, the product in \eqref{eq:Qmn-general} runs over \(\mathcal O(L)\) terms and will vanish  unless the escape\mbox{‑}rate ratios \(\Omega_{\bar k}/\Omega_k\) decays  sufficiently fast with system size. For example, if \(\Omega_{\bar k}/\Omega_k=\mathrm{const}>0\), then \(\prod_{k<n}\big(1-\Omega_{\bar k}/\Omega_k\big)\to 0\) and the largest weight of surviving systems occur at $n=1$ (or at density $\rho=1/L$). With  increase of  $L$  this  weight increases  in expense of weights for $n>1$   and   in the thermodynamic limit  the QS distribution   becomes  a Dirac-delta function  $\delta(\rho).$
By contrast, sufficiently fast decay of these ratios with \(L\) (e.g., exponentially in \(L\), as in our BDD class)  can yield non-vanishing  
 QS weights at some $\rho=\rho^*\ne 0,$  that depends on the initial  condition. This   mechanism is responsible for   the   non unique QS behavior observed in  BDD   models in \autoref{sec:NoBirth}.
 
Based on our findings, we conjecture that the long‑time survival behavior in a Markov process is nonunique (i.e., it retains memory of the initial open CC) if and only if:
\begin{enumerate}
  \item the nonabsorbing part splits into at least two macroscopic open CCs; and
  \item if the open CCs have the topology of a linear directed acyclic graph terminating at a unique absorbing sink $\mathcal S_0$ (i.e., \(\mathcal S_L \to \mathcal S_{L-1} \to \dots \to \mathcal S_1 \to \mathcal S_0\)) with escape rates  $\Omega_k$ from $\mathcal S_k\to \mathcal S_{k-1}$, then for nonuniqueness  to occur,  the ratios of  the slowest 
 escape‑rate   say $\Omega_{\bar k}$ to  all  other  rates   for $k<\bar k$  must  vanish  in the thermodynamic limit as $1/L$ or faster, i.e, 
  \begin{equation}\label{eq:conjectureEq}
    \frac{\Omega_{\bar k}}{\Omega_{k}} \;\sim\; \frac{1}{L^{u}},
  \qquad u\ge 1,
  \qquad \forall\,k < \bar k.    
  \end{equation}
\end{enumerate}
Equivalently, this vanishing is the minimal structural requirement for conditional long‑time weights of the form in Eq.~\eqref{eq:Qmn-general} to retain memory of the initial condition in the thermodynamic limit. If the ratios \(\Omega_{\bar k}/\Omega_k\) decays slower than $1/L$, the long‑time behavior is unique, and independent of the initial condition. Note, that  an  exponentially  slow rate, that  we demonstrate in \autoref{sec:NoBirth}, is  not an  essential requirement for nonuniqueness - other choices, say   \(\Omega_{\bar k}/\Omega_k \sim 1/L\) can also lead to a non-unique QS state.  This  bottleneck  scenario  is discussed further  in Appendix~\ref{AppendixSec:Bottleneck}.

\section{Discussion and conclusion}
We  have identified possible  structure  in  Markov processes with absorbing states that  leads to nonunique thermodynamic behavior
and presented it into a precise conjecture for thermodynamic nonuniqueness: {\it nonuniqueness occurs if and only if the active configuration space fractures into multiple macroscopic open communicating classes (CCs) and the escape‑rate ratios between the slowest class and all competitors vanish with system size (minimally as an inverse power of \(L\)).}

\begin{figure}[!t]
\includegraphics[width=\columnwidth]{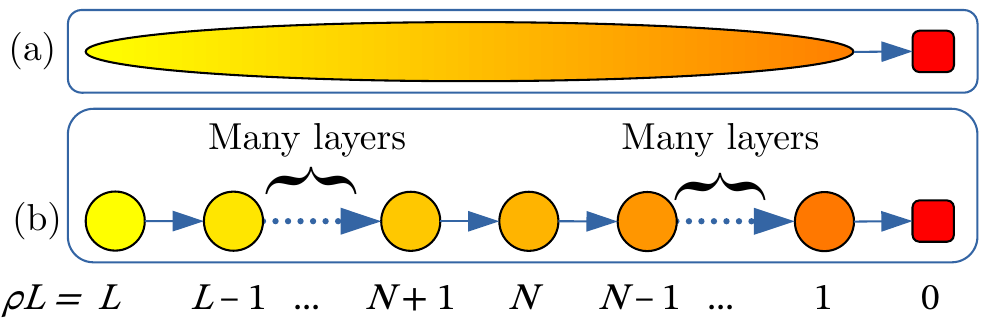}
\caption{
  \textbf{Schematic communicating-class structure in birth–death–diffusion models.} Circles denote open CCs and squares closed (absorbing) CCs; arrows indicate allowed transitions. Classes are ordered by decreasing order parameter \(\rho\).
  (a) With birth active, the nonabsorbing sector collapses into a single open CC, yielding a unique QS state.
  (b) No-birth limit with transitions among open CCs.
  (c) No-birth limit with multiple open CCs and inter-species transitions.
}
\label{fig:CC}
\end{figure}

We demonstrated the conjecture in a birth–death–diffusion (BDD) class with fast mixing and slow, density‑dependent reactions. The time‑scale of separation enables an exact reduction to an effective one‑dimensional dynamics in particle number, from which the phase diagram and long‑time behavior follow. With birth allowed, the active sector forms a single open CC (Fig.~\ref{fig:CC}(a)) and the long‑time behavior—conditioned on survival—is unique. Turning birth off splits the active sector into fixed‑particle‑number open CCs (Fig.~\ref{fig:CC}(b)), and 
the restricted generator becomes reducible; the resulting long‑time behavior retains memory of the initial sector and, at absorbing criticality, fluctuation scaling can depend explicitly on the initial condition. 
When the initial density follows a power-law distribution with exponent $a$, Eq.~\eqref{eq:init_a}, the critical exponents of the absorbing phase transition acquire a continuous dependence on $a$, as shown in Eq.~\eqref{eq:gamma}.

Continuous variation of critical exponents is well known in systems possessing marginal parameters, both theoretically~\cite{Kosterlitz1974,Baxter1971,Pearce1987,Indranil2023} and with experimental indications~\cite{Suzuki1992,Khan2017}. In disordered systems, numerical simulations have sometimes reported apparent violations of universality~\cite{Hartmann2001,Malakis2009}; however, several such cases were later shown to belong to a single universality class once strong corrections to scaling were properly accounted for~\cite{Fytas2013}. In contrast, an explicit dependence of critical exponents on the initial condition constitutes a defining characteristic of the BDD class of models at the absorbing phase transition. This behavior arises from the fractured structure of the active sector and the vanishing ratios of escape rates, leading to a distinct form of memory in the critical dynamics.

The implications are far-reaching:  it poses a direct challenge to the conventional notion of universality, revealing a non-equilibrium phase transition whose critical exponents  carry the memory of the  
initial  condition  and vary continuously  as  the parameters in  the initial  condition change. This finding in BDD class of  systems suggests a fundamental departure from the established framework of critical phenomena  where 
the universality hypothesis based on  renormalization group theory erases  the microscopic  details  and produce a 
unique set of critical exponents as long as interaction is local  and consistent with the symmetry.  The  dependence of the critical exponents on initial condition  is not   an outcome  of  the  non-uniqueness of the steady state alone, the topological structure  of the Markovian dynamics and choice of transition rates matter. Indeed,  in this article we also present an explicit counterexample  (Appendix \ref{AppendixSec:Bottleneck})   where  the critical exponents remain invariant even when  the steady state is not unique.

The  thermodynamic nonuniqueness  conjecture   that we propose is broadly applicable beyond the BDD realization. Many stochastic systems naturally organize into multiple open CCs such as such as ecological metapopulations with local extinction \cite{Hanski1999,Ovaskainen2010}; epidemiological compartment models with absorbing disease–free states  \cite{AndersonMay1991,KeelingRohani2008,Nasell1999,Nasell2001}; and biochemical reaction networks \cite{Feinberg2019,AndersonKurtz2015}. In all these settings, long‑lived active dynamics can retain memory of preparation, and the conjecture predicts when this memory survives in the thermodynamic limit and when it can imprint on critical scaling. The structural nature of the criterion makes it testable across conservative dynamics and dimensions.
  
Taken together, our results revise a common assumption in absorbing‑state transitions, namely, uniqueness of long‑time behavior conditioned on survival, and point to a class of systems where history matters in a measurable way, including at criticality.

\section*{Acknowledgement}
We thank Gianluca Teza for discussions. P.K.M. acknowledges the financial support provided by ANRF, Science and Engineering Research Board (SERB), DST, Government of India, under Grant No. MTR/2023/000644. P.K.M. also acknowledges MPIPKS for supporting his visit, during which part of this work was carried out.

\appendix

\section{Communicating classes in Markov processes}
\label{AppendixSec:CommunicatingClasses}

In a Markov process with transition matrix $\mathbf T$ (entries $\mathbf T_{ij}$ denote the transition rate from state $i$ to $j$, with $\mathbf T_{ii}=0$), a state $m$ is accessible from a state $k$ if $\exists$ a path from $k$ to $m$, i.e., $(\mathbf T^{\,n})_{km}>0$ for some $n\ge 0$; we write $k\to m$. If both $k\to m$ and $m\to k$, then $k$ and $m$ communicate, denoted $k\leftrightarrow m$. The communicating classes (CCs) are the equivalence classes of states under this relation: two states are in the same CC {\it iff} (if and only if) they communicate \cite{Norris1998,Levin2017}. Communication is an equivalence relation (reflexive, symmetric, transitive); for example, if $k\leftrightarrow m$ and $m\leftrightarrow \ell$, then $k\leftrightarrow \ell$.
For example, for the Markov dynamics on the state space $S=\{0,1,2,3,4\}$ illustrated below,
\begin{figure}[h]
\includegraphics[width=0.4\columnwidth]{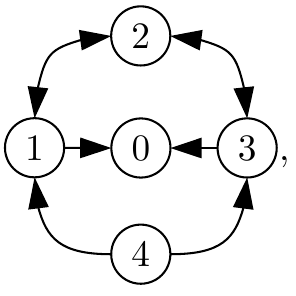}
\end{figure}

\noindent the CCs are $\{0\}$, $\{1,2,3\}$, and $\{4\}$. A CC is \emph{closed} if it cannot be left; here $\{0\}$ is closed. CCs that are not closed are \emph{open}; here $\{1,2,3\}$ and $\{4\}$ are open.

\section{Communicating‑class structure in finite‑state Markov chains governs the uniqueness of long‑time behavior}
\label{AppendixSec:MultipleOpenCCs}
In the introduction, we illustrated with help of an example with finite number of states that when the nonabsorbing sector of a Markov process consists of a single open communicating class (CC), the long‑time survival behavior is unique (i.e., independent of the initial condition). By contrast, when the nonabsorbing sector fractures into multiple open CCs, the long‑time survival behavior can depend on initial conditions, i.e., open CC the initial state belongs to.

We considered a simple nine‑state continuous‑time Markov process with state space \(S=\{1,2,\dots,9\}\) in which state \(9\) is absorbing (Fig.~\ref{Fig:FigExplainCC_alternative}). We use the column‑sum–zero convention for generators: \(M_{i,j}\ge 0\) for \(i\neq j\) is the rate of transitions \(j\to i\), and each column of \(\mathbf M\) sums to zero. The evolution of the probability vector \(|P(t)\rangle=(P_1(t),\dots,P_9(t))^{T}\) is governed 
\begin{equation}
\frac{d}{dt}\,|P(t)\rangle = \mathbf M\,|P(t)\rangle,
\end{equation}
with the following choice of rate matrix
\begin{equation}
\mathbf M = \begin{pmatrix}
 -0.1 & 1 & 0 & 1 & 0 & 0 & 0 & 0 & 0 \\
  0.1 & -3 & 1 & 0 & 0 & 0 & 0 & 0 & 0 \\
  0 & 1 & -2 & 0 & 0 & 0 & 0 & 0 & 0 \\
  0 & 0 & 0 & -3 & 1 & 0 & 0.8 & 0 & 0 \\
  0 & 1 & 0 & 0 & -2 & 0.3 & 0 & 0 & 0 \\
  0 & 0 & 1 & 0 & 1 & -0.3 & 0 & 0 & 0 \\
  0 & 0 & 0 & 2 & 0 & 0 & -1.8 & 1 & 0 \\
  0 & 0 & 0 & 0 & 0 & 0 & 1 & -2 & 0 \\
  0 & 0 & 0 & 0 & 0 & 0 & 0 & 1 & 0 \\
\end{pmatrix}.
\end{equation}

We plot the time evolution of the average survival states (states 1 to 8) in Fig.~\ref{Fig:FigExplainCC_alternative} (b), defined as
\begin{equation}
\langle n(t)\rangle = \frac{1}{\mathbb S(t)} \sum_{i=1}^{8} i\,P_i(t),
\end{equation}
where $\mathbb S(t) = \sum\limits_{i=1}^{8} P_i(t)$ is the survival probability at time \(t\) (i.e., the probability of not being in the absorbing state). For the matrix above, the nonabsorbing sector \(S^+=\{1,\dots,8\}\) is a single open CC, and the conditional evolution is irreducible. Consistent with the Perron–Frobenius theorem, we find that \(\langle n(t)\rangle\) converges to the same asymptotic value for different initial conditions, e.g., starting from \(|P(0)\rangle=(1,0,0,0,0,0,0,0,0)^T\) (all probability in state 1) or \(|P(0)\rangle=(0,0,0,1,0,0,0,0,0)^T\) (all probability in state 4). See Fig.~\ref{Fig:FigExplainCC_alternative}(a, b).

Now delete the transition \(4\to 1\) by setting \(M_{1,4}=0\), and consequently the total out rate from state $4$ is $\mathbf M_{4,4} = -2$. This single modification splits the active sector into multiple open CCs (see Fig.~\ref{Fig:FigExplainCC_alternative}(c)) and the restricted generator on \(S^+\) becomes reducible. In this case, \(\langle n(t)\rangle\) converges to distinct asymptotic values depending on the initial condition, e.g., initializing in state \(1\) versus state \(4\) yields different long‑time limits, demonstrating explicit initial‑condition dependence in a finite system; see Fig.~\ref{Fig:FigExplainCC_alternative}(d).

More generally, uniqueness of the quasi‑stationary (QS) state can be assessed by the following steps:
\begin{enumerate}
  \item Partition the state space into communicating classes (CCs).
  \item Classify each CC as open or closed. If there is exactly one macroscopic open CC, the QS state is unique.
  \item If there are multiple open CCs, compute the escape (inter‑class) rates \(\Omega_k\) from each open class \(\mathcal S_k\) to downstream classes (including toward the absorbing class). Nonunique QS behavior (memory of the initial class) can arise when a slow open class not directly connected to the absorbing class has an escape rate asymptotically smaller than all competitors.
\end{enumerate}

\section{Why quasi-stationary states are generally unique}
\label{AppendixSec:UniquenessGeneral}
\begin{figure}[!htbp]
  \centering
  \includegraphics[width=\columnwidth]{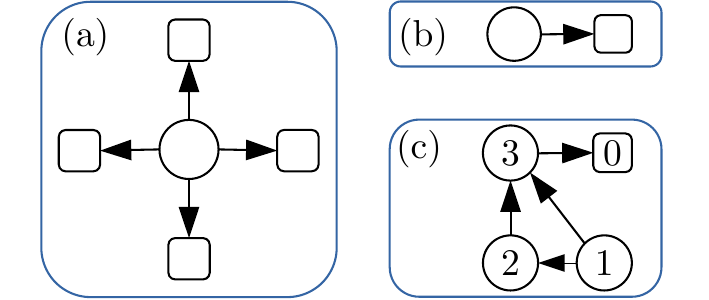}
  \caption{
  \textbf{Schematic communicating-class structure in various models.} A system with many closed CCs (squares) and one open CC (discs).
  (a) Examples include the pair contact process \cite{Jensen1993} and branching processes; in all these models, each closed communicating class contains only one unique absorbing configuration. 
  (b) Systems with one open and one closed CC; most models with a single absorbing configuration, such as directed percolation \cite{Kinzel1983, Hinrichsen2000} and the contact process \cite{Harris1974} in any dimension. 
  (c) Systems with many open CCs and one closed CC.}
  \label{fig:ccexamples}
\end{figure}

Based on the definition of communicating classes (Appendix~\ref{AppendixSec:CommunicatingClasses}), typical CC structures across models are shown in Fig.~\ref{fig:ccexamples}: (a) the pair contact process \cite{Jensen1993} and branching processes; (b) directed percolation \cite{Kinzel1983, Hinrichsen2000} and the contact process \cite{Harris1974} in any dimension.

Many models with a single absorbing configuration, such as directed percolation \cite{Kinzel1983,Hinrichsen2000} and the contact process \cite{Harris1974}, have one open CC (the nonabsorbing sector $S^+$) and one closed CC (the absorbing configuration $\mathcal C_0$), see Fig. \ref{fig:ccexamples} (b). Let $w_{\mathcal C_i\to\mathcal C_j}\!\ge 0$ denote the transition rate from configuration $\mathcal C_i$ to $\mathcal C_j$. Writing the probability vector as $|P(t)\rangle$, the master equation has the block form
\begin{equation}
\frac{d}{dt}\,|P(t)\rangle
= \mathbf M\,|P(t)\rangle,
\qquad
\mathbf M=\begin{pmatrix}
0 & \langle w|\\
\mathbf 0 & \mathbf A
\end{pmatrix},
\label{eq:2x2}
\end{equation}
where $\langle w|=\big(w_{\mathcal C_1\to\mathcal C_0},w_{\mathcal C_2\to\mathcal C_0},\dots\big)$ collects the escape rates to the absorbing state and $\mathbf A$ is the generator restricted to the open CC, with
$\mathbf A_{ij}=\mathbf C_{ij}-\delta_{ij}w_{\mathcal C_i\to\mathcal C_0}$ and $\mathbf C$ the (irreducible) dynamics within $S^+$. If $\lambda_i$ are the eigenvalues of $\mathbf A$ ordered by decreasing real part, and $\langle\phi_i|,|\phi_i\rangle$ are the corresponding left and right eigenvectors, then the eigenvalues of $\mathbf M$ are $\{0\}\cup\{\lambda_i\}$ and the corresponding left and right eigenvectors are
\begin{eqnarray} 
\langle \Psi_0| = (1,1,\dots),\quad |\Psi_0 \rangle = (1,0,0,\dots)^T,\nonumber\\
\langle \Psi_{i\neq 0}| = \begin{pmatrix}0,&\langle \phi_i |\end{pmatrix},\quad |\Psi_{i\neq 0} \rangle = \begin{pmatrix} \dfrac {\langle w | \phi_i \rangle} {\lambda_i}\\ |\phi_i\rangle \end{pmatrix}.
\end{eqnarray}

The steady state $|\Psi_0\rangle = \begin{pmatrix} 1 & 0 & 0 & \dots\end{pmatrix}^T$ of the system is trivial here: in the limit $t\to\infty$ the system is certainly absorbed. The quasi-stationary probability $\mathbf{Q}_{ij}$ is the probability that, starting from a configuration $\mathcal C_{i\ne0}$, the system survives and is found in configuration $\mathcal C_{j\ne 0}$:
\begin{equation}
\mathbf{Q}_{ij} = \lim_{t\to\infty}  \frac{\sum\limits_{n}\langle j|e^{\lambda_n t}|\Psi_n\rangle\langle \Psi_n|i\rangle}{\sum\limits_{k\neq 0,n}\langle k|e^{\lambda_n t}|\Psi_n\rangle\langle \Psi_n|i\rangle}
= \frac{\langle j|\Psi_1\rangle}{ \sum\limits_{k\ne0} \langle k|\Psi_1\rangle }.
\end{equation}
Irreducibility of the dynamics on $S^+$ ensures positivity of $\langle\phi_1|$ and $|\phi_1\rangle$, the left and right eigenvectors corresponding to the largest eigenvalue of $\mathbf A$ \cite{Keizer1972, Hyver1972}. The positivity of $\langle\phi_1|$ implies $\langle\Psi_1|i\neq 0\rangle>0$. Thus, the quasi-stationary measure is the normalized right eigenvector associated with $\lambda_1$, the largest eigenvalue of $\mathbf A$ (equivalently, the second-largest eigenvalue of $\mathbf M$), and it does not depend on the initial condition.

For systems with many absorbing configurations $S^0=\{\mathcal C_{0,1},\mathcal C_{0,2},\dots\}$, such as the pair contact process \cite{Jensen1993}, there are many closed CCs and one open CC as in Fig. \ref{fig:ccexamples} (a). The dynamics is given by the master equation
\begin{equation}
\frac {d}{d t}|P(t)\rangle = \mathbf M|P(t)\rangle,\quad
\mathrm{with}\,{\bf M}=\begin{pmatrix}
     \mathbf 0 & \mathbf w\\
     \mathbf 0 & {\bf A}
    \end{pmatrix},
\end{equation}
where the $\mathbf 0$ in the upper-left block is a square null matrix of dimension equal to the number of absorbing configurations, $\mathrm n(S^0)$. A calculation analogous to the single-absorbing-state case leads again to a unique quasi-stationary state: irreducibility within the nonabsorbing subspace $S^+=S/S^0$ guarantees uniqueness of the QS distribution.

When there are many open CCs, the dynamics within the nonabsorbing subspace $S^+$ is no longer irreducible and this can lead to a nonunique quasi-stationary state. Consider, for example, Fig. \ref{fig:ccexamples} (c), with three open CCs labeled $k=1,2,3$, each containing $n_k$ configurations, and one closed CC labeled $k=0$ with $n_0=1$ (an absorbing configuration). Then the matrix $\mathbf M$ in the master equation is
\begin{equation}
{\mathbf  M}= \begin{pmatrix} 
0& {\mathbf  0}& {\mathbf  0}&   \langle w_3|\\
{\mathbf  0}&  {\mathbf  A_{11}} &{\mathbf  0}& {\mathbf  0}\\
{\mathbf  0}&  {\mathbf  A_{21}} &{\mathbf  A_{22}}& {\mathbf  0}\\
{\mathbf  0}&  {\mathbf  A_{31}} &{\mathbf  A_{32}}& {\mathbf  A_{33}}\\
\end{pmatrix},
\end{equation}
where $\mathbf A_{kk'}$ are matrices of dimension $n_k\times n_{k'}$. Writing ${\mathbf M}$ as a $2\times 2$ block matrix as in Eq.~(\ref{eq:2x2}), we have $\langle w| = (\mathbf 0,\,\mathbf 0,\, \langle w_3|)$ and
\begin{equation} 
 {\mathbf A}=  \begin{pmatrix} 
 {\mathbf  A_{11}} &{\mathbf  0}& {\mathbf  0}\\
 {\mathbf  A_{21}} &{\mathbf  A_{22}}& {\mathbf  0}\\
 {\mathbf  A_{31}} &{\mathbf  A_{32}}& {\mathbf  A_{33}}\\
\end{pmatrix},
\end{equation}
which is reducible. It is evident that the presence of more than one open communicating class renders $\mathbf A$ reducible and may break the positivity condition of the leading left and right eigenvectors of largest eigenvalue $\lambda$, thereby allowing the quasi-stationary state to be nonunique.

On the other hand, physical systems like sand-pile models \cite{Bak1987,Dhar1990,Dhar2006,Chhajed2021c} exhibit a unique quasi-stationary state for the following reason. Sandpile models are driven dissipative systems originally introduced to explain self‑organized criticality (SOC) \cite{Bak1987}. Grains are added; sites whose occupation exceeds a local threshold topple and redistribute grains; boundary dissipation continues until all sites are stable. Fixed‑energy sandpiles (FES) \cite{Dickman2000,Basu2012,Park2018} remove drive and dissipation so the total grain number is conserved; tuning the density produces an absorbing‑state transition between a persistently active regime and one of exponentially many absorbing configurations. At subcritical densities the configuration space contains many absorbing (closed) communicating classes. Above threshold the active (nonabsorbing) sector can itself fracture into multiple open communicating classes distinguished by a conserved imbalance of grains in even–odd (sublattice) under synchronous toppling: starting from an imbalance, each unstable odd site topples to even neighbors (and vice versa), so the parity pattern is preserved and mixing between classes with different even-odd grain imbalance are rare. Introducing stochasticity in local thresholds (Oslo sandpile model \cite{Frette1993,Christensen1996,Grassberger2016}) adds transition channels between the imbalance classes and hence recovers uniqueness of the quasi‑stationary state.

\section{Nonunique quasi‑stationary states with bottlenecked escape rates}
\label{AppendixSec:Bottleneck}
Consider a reducible nonabsorbing sector \(S^+=\bigcup_{n=1}^{L}\mathcal S_n\) in which typical sectors \(\mathcal S_k\) have \(\mathcal O(L)\) outflux channels at unit rate, so \(\Omega_k\sim L\), while a bottleneck sector \(\mathcal S_{dL}\) has only \(\mathcal O(1)\) outflux channels, so \(\Omega_{dL}\sim 1\). Then
\begin{equation}
    \frac{\Omega_{dL}}{\Omega_k} \sim \frac{1}{L}\qquad (k\neq dL),
\end{equation}
and the product \(\prod_{k<\bar k}\big(1-\Omega_{\bar k}/\Omega_k\big)\) over \(\mathcal O(L)\) terms approaches a finite, nonzero limit as \(L\to\infty\). In this case the contributing slow mode set minimizes \(\Omega_k\) among those overlapping the initial sector, and one finds, in the continuum,
\begin{equation}
Q(\rho_{\mathrm{in}},\rho)=\Theta(d-\rho_{\mathrm{in}})\,\delta(\rho-d),
\label{eq:QmnContinuum-bottleneck}
\end{equation}
where $\rho_{\mathrm{in}}=i/L$ is initial density. The ensemble QS distribution is
\begin{equation}
\mathbb{Q}(\rho)=\frac{1}{\mathcal N}\int_0^{1}\mathrm d\rho_{\mathrm{in}}\, Q(\rho_{\mathrm{in}},\rho)\,\pi(\rho_{\mathrm{in}}),
\end{equation}
normalized by \(\mathcal N\). The absorbing order parameter and its variance scale as
\begin{equation}
\varrho = \Delta^{\beta},\quad \beta=1,\quad
\langle\rho^2\rangle-\langle\rho\rangle^2 =0,\quad \gamma=0,
\end{equation}
with \(\Delta=d-d_c\) and \(d_c=0\). Thus, this bottleneck mechanism yields nonunique QS states while the APT exponents \(\beta,\gamma\) are independent of \(\pi(\rho_{\mathrm{in}})\), in contrast to the exponential‑ratio BDD class where susceptibility exponents can depend on the initial distribution.
\bibliographystyle{unsrt}
\bibliography{mybib}

\end{document}